\documentclass{PoS}

\usepackage{amsfonts} 
\usepackage{amssymb} 
\usepackage{amsmath} 
\usepackage{subfigure}
\usepackage{graphicx} 
\usepackage{array} 
\usepackage{dcolumn} 
\usepackage{bm} 
\usepackage{latexsym} 
\usepackage{longtable} 
\usepackage{color}
\usepackage{mathrsfs}

\graphicspath{{./figs/}}

\newcommand{\CO}{{\cal O}}

\newcommand{\Tr}{\textrm{Tr}}

\newif\ifContLineOne
\newif\ifContLineTwo
\newif\ifContLineThree

\def\conC#1{\vbox{\ialign{##\crcr
  \ifContLineThree\hrulefill\else\vphantom{\hrulefill}\fi\crcr
  \noalign{\kern3.2pt\nointerlineskip}
  \ifContLineTwo\hrulefill\else\vphantom{\hrulefill}\fi\crcr
  \noalign{\kern3.2pt\nointerlineskip}
  \ifContLineOne\hrulefill\else\vphantom{\hrulefill}\fi\crcr
  \noalign{\nointerlineskip}
  $\hfil\textstyle{\vbox to 14pt{}#1}\hfil$\crcr}}}

\def\DrawLeg#1#2{
  \kern-.2pt              
  \dimen2 =#1             
  \advance\dimen2 by 2pt  
  \dimen3 = 10.6pt        
  \dimen4 =3.6pt          
  \advance\dimen3 by -\dimen2
  \multiply\dimen4 by #2
  \advance\dimen3 by \dimen4
  \raise\dimen2 \hbox{\vrule height\dimen3 width .4pt} 
  \kern-.2pt}             

\def\begC#1#2{\setbox0 =\hbox{$\textstyle{#2}$}
  \dimen0=.5\wd0 \dimen1=\ht0
  \conC{\hskip\dimen0}
  \count255=#1
  \ifnum\count255 =1 \ContLineOnetrue\else
  \ifnum\count255 =2 \ContLineTwotrue\else
  \ifnum\count255 =3 \ContLineThreetrue\fi\fi\fi
  \DrawLeg{\dimen1}{\count255}
  \conC{\hskip\dimen0}
  \kern-\dimen0\kern-\dimen0 \box0}

\def\endC#1#2{\setbox0 =\hbox{$\textstyle{#2}$}
  \dimen0=.5\wd0 \dimen1=\ht0
  \conC{\hskip\dimen0}
  \count255=#1
  \ifnum\count255 =1 \ContLineOnefalse\else
  \ifnum\count255 =2 \ContLineTwofalse\else
  \ifnum\count255 =3 \ContLineThreefalse\fi\fi\fi
  \DrawLeg{\dimen1}{\count255}
  \conC{\hskip\dimen0}
  \kern-\dimen0\kern-\dimen0 \box0}

\title{ 
  Taste non-Goldstone pion decay constants in staggered chiral perturbation theory 
}

\ShortTitle{
  Taste non-Goldstone pion decay constants in SChPT 
}

\author{Jon A.~Bailey, \speaker{Boram Yoon}, Weonjong Lee\\
  Lattice Gauge Theory Research Center, CTP, and FPRD, \\
  Department of Physics and Astronomy,
  Seoul National University, Seoul, 151-747, South Korea \\
  E-mail: \email{boramsnu@gmail.com}}

\author{SWME Collaboration}

\abstract{
  We calculate the next-to-leading order axial current decay constants of 
  taste non-Goldstone pions and kaons in staggered chiral perturbation theory.
  This is an extension of the taste Goldstone decay constants 
  calculation to that of the non-Goldstone tastes.
  We present results for the partially quenched case in the SU(3) and SU(2) 
  staggered chiral perturbation theories
  and discuss the difference between the taste Goldstone and non-Goldstone
  cases.
  }

\FullConference{
The 30th International Symposium on Lattice Field Theory - Lattice 2012 \\
June 24 - 29, 2012\\
Cairns, Australia
}

\begin{document}

\section{Introduction} 

Staggered chiral perturbation theory (SChPT) was developed to describe
lattice data generated by staggered fermions, which have an exact chiral
symmetry at nonzero lattice spacing.
Using SChPT, lattice results can be extrapolated to the physical
quark masses and the continuum limit, removing dominant lattice artifacts
coming from the taste symmetry breaking of staggered fermions.

In Ref.~\cite{Aubin:2003uc}, Aubin and Bernard calculated next-to-leading order
(NLO) corrections to the decay constants of taste Goldstone pions
and kaons, associated with the exact chiral symmetry of the staggered action, 
in SChPT.
Here we extend the calculation to the taste non-Goldstone pions and kaons. 

In Sec.~\ref{sec:review}, we consider the leading order (LO) and NLO terms of 
the chiral Lagrangian that contribute to the decay constants.
In Sec.~\ref{sec:decay_consts}, we outline the calculation of NLO corrections
to the decay constants of taste non-Goldstone pions and kaons,
and write results in a theory with three flavors and four tastes for each flavor.
In Sec.~\ref{sec:results}, we present the results for the partially quenched 
case in the SU(3) and SU(2) SChPT,
and we conclude in Sec.~\ref{sec:conclusion}.
Unless defined explicitly we use the notation of Ref.~\cite{SWME:2011aa}. 

\section{\label{sec:review}Chiral Lagrangian for staggered quarks}
The chiral Lagrangian for staggered quarks was formulated by Lee and Sharpe
for the single-flavor case \cite{Lee:1999zxa} and generalized by Aubin and 
Bernard to multiple flavors \cite{Aubin:2003mg}.
In the standard power counting,
\begin{equation}
 \CO(p^2/\Lambda_\chi^2)
 \approx \CO(m_q/\Lambda_\chi)
 \approx \CO(a^2\Lambda_\chi^2),
\label{eq:count}
\end{equation}
the order of a Lagrangian operator is the sum of non-negative integers, 
$n_{p^2}$, $n_m$ and $n_{a^2}$, which are the number of derivative
pairs, number of quark mass factors, and powers of the squared lattice spacing
in the operator, respectively.
At LO, the Lagrangian operators fall into three classes:
$(n_{p^2}, n_m, n_{a^2}) = (1,0,0)$, $(0,1,0)$ and $(0,0,1)$,
and we have
\begin{align}
 \label{F3LSLag}
 \mathcal{L}_\mathrm{LO} =
  &\frac{f^2}{8} \Tr(\partial_{\mu}\Sigma \partial_{\mu}\Sigma^{\dagger}) - 
    \frac{1}{4}\mu f^2 \Tr(M\Sigma+M\Sigma^{\dagger})
  + \frac{2m_0^2}{3}(U_I + D_I + S_I)^2 + a^2 (\mathcal{U+U^\prime})
\,,
\end{align}
where $f$ is the decay constant at LO, $\mu$ is a constant in
the unit of mass, $M$ is the mass matrix, $\Sigma \equiv \exp(i\phi/f)$, 
and $\phi$ is the pseudo-Goldstone boson (PGB) field.
The term multiplied by $m_0^2$ is the anomaly contribution,
and $\mathcal{U}$ and $\mathcal{U^\prime}$ are the taste symmetry 
breaking potentials defined in Ref.~\cite{Aubin:2003mg}.

At NLO, there are six classes that satisfy 
$n_{p^2} + n_m + n_{a^2} = 2$.
Operators in two classes contribute to the decay constants: 
$(n_{p^2}, n_m, n_{a^2}) = $ $(1,1,0)$ and $(1,0,1)$.
The contributing operators in the class $(1,1,0)$ 
are Gasser-Leutwyler terms \cite{Gasser:1984gg},
\begin{align}
\label{eq:GLterms}
\mathcal{L}_\mathrm{GL}
 = L_4\mathrm{Tr}(\partial_\mu\Sigma^\dagger\partial_\mu\Sigma)
       \mathrm{Tr}(\chi^\dagger\Sigma+\chi\Sigma^\dagger)
  + L_5\mathrm{Tr}(\partial_\mu\Sigma^\dagger\partial_\mu\Sigma
       (\chi^\dagger\Sigma+\Sigma^\dagger\chi))
\,,
\end{align}
where $L_4$ and $L_5$ are low-energy constants (LECs) and 
$\chi = 2\mu M$.
The contributing operators in the class $(1,0,1)$ 
are terms given by Sharpe and Van de Water in Ref.~\cite{Sharpe:2004is}.

\section{\label{sec:decay_consts}Decay constants of flavor-charged
  pseudo-Goldstone bosons}
The decay constant $f_{P_t^+}$ for a flavor-charged PGB $P_t^+$ with 
taste $t$ is defined by the matrix elements
\begin{equation}
\label{eq:def_decay_const}
 \langle 0 | j_{\mu 5, t}^{P^+} | P_t^+(p) \rangle  = -i f_{P_t^+} p_\mu,
\end{equation}
where $j_{\mu 5, t}^{P^+}$ is the axial current.
From the LO Lagrangian, the LO axial current is
\begin{equation}
 \label{eq:axial_curr}
 j_{\mu 5, t}^{P^+} 
  =  -i\frac{f^2}{8} \Tr 
   \left[ T^{t(3)} \mathcal{P}^{P^+} 
    (\partial_\mu \Sigma \Sigma^\dag + \Sigma^\dag \partial_\mu \Sigma)
   \right],
\end{equation}
where $T^{a(3)} \equiv I_3 \otimes T^a$, $I_3$ is the identity matrix in flavor
space, and $\mathcal{P}^{P^+}$ is a projection operator that chooses $P^+$
from the $\Sigma$ field, defined by $\mathcal{P}^{P^+}_{ij} = \delta_{ix}\delta_{jy}$.
For flavor-charged states, which are the interest in this paper, $x \neq y$.
Note that $\Sigma = \exp(i\phi/f)$ can be expanded in terms of $\phi$.

There are three types of NLO corrections to the decay constants:
(a) one-loop wavefunction renormalization correction from the $\mathcal{O}(\phi)$ term 
of the axial current, $\delta f^Z_{P^+_t}$,
(b) one-loop correction from the $\mathcal{O}(\phi^3)$ term 
of the axial current, $\delta f^\textrm{current}_{P^+_t}$, and
(c) analytic contribution from the NLO terms of the axial current and the analytic
contribution to the self-energy,
$\delta f^\textrm{anal}_{P^+_t}$.
Combining these three types of corrections (a) -- (c), we write the decay constants:
\begin{equation}
 f_{P_t^+} = f\left[ 1 + \frac{1}{16\pi^2 f^2} 
  \left( 
    \delta f^Z_{P^+_t} + \delta f^\textrm{current}_{P^+_t} \right)
    + \delta f^\textrm{anal}_{P^+_t}
  \right]\label{eq:ftot}.
\end{equation}
\begin{figure}[t]
\centering
  \subfigure[]{
    \label{fig:waveftn_crxn}
    \includegraphics[width=8pc]{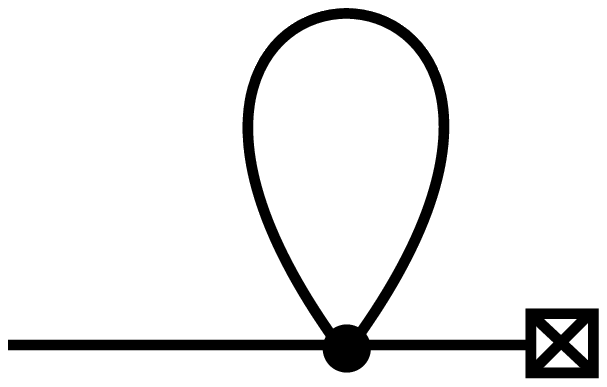}
  }
\qquad
  \subfigure[]{
    \label{fig:current_crxn}
    \includegraphics[width=6.2pc]{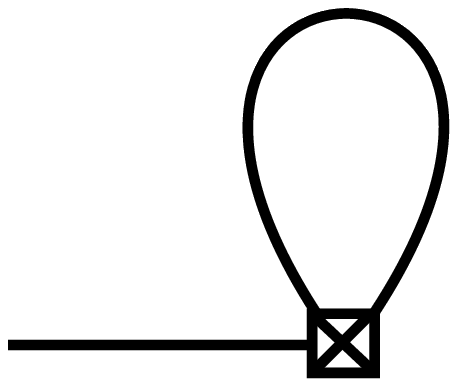}
  }
\caption{Diagrams contributing to the decay constants.
  (a) is the wavefunction renormalization correction and
  (b) is the current correction.}
\end{figure}

First we consider the wavefunction renormalization correction.
Considering the $\mathcal{O}(\phi)$ term of the LO axial current,
$j_{\mu 5, t}^{P^+, \phi}  = f \left( \partial_\mu \phi_{yx}^t \right)$,
we find contributions to the decay constants,
\begin{align}
 \langle 0 | j_{\mu 5, t}^{P^+, \phi} | P_t^+(p) \rangle
  = f (-ip_\mu) \langle 0 | \phi_{yx}^t | P_t^+(p) \rangle 
  = f (-ip_\mu) \sqrt{Z_{P_t^+}}
\,.
\end{align}
Here $Z_{P_t^+} \equiv 1 + \delta Z_{P_t^+}$ is the wavefunction
renormalization constant of the $\phi$ field.
$\delta Z_{P_t^+}$ gives the NLO correction to the decay constants,
\begin{equation}
 \delta f^Z_{P_t^+} 
  \equiv \frac{16\pi^2 f^2}{2}\delta Z_{P_t^+}
  = - \frac{16\pi^2 f^2}{2} \frac{d\Sigma(p^2)}{dp^2},
\end{equation}
where $\Sigma(p^2)$ is the self-energy of $P_t^+$.
Using the self-energy from Ref.~\cite{SWME:2011aa}, we find the
one-loop corrections
\begin{align}
\label{eq:delta_f_z}
 \delta f
 ^Z_{P_t^+} 
  = \frac{1}{24} \sum_a 
   \Bigg[ 
    \sum_{Q} l(Q_a) 
  + 16\pi^2 \int \frac{d^4 q}{(2\pi)^4} 
     \left(
      D_{xx}^a + D_{yy}^a - 2\theta^{at} D_{xy}^a
     \right)
   \Bigg],
\end{align}
where $l(Q_a)$ is the chiral logarithm and $D^a_{ij}$ is the disconnected piece 
of the propagator.
Here $Q$ runs over six flavor combinations, $xi$ and $yi$ for $i\in\{u,d,s\}$,
$a$ runs over the 16 PGB tastes in the $\mathbf{15}$ and $\mathbf{1}$ of $SU(4)_T$,
and $Q_a$ is the squared tree-level meson mass with flavor $Q$ and taste $a$.
The self-energy also contains NLO analytic corrections to the decay
constants, which we discuss below.

Next we consider the loop correction from the $\mathcal{O}(\phi^3)$ terms 
of the LO axial current,
\begin{align}
\label{eq:axial_curr_phi3}
 j_{\mu 5, t}^{P^+, \phi^3}
 = - \frac{1}{24f} \tau_{tabc}  \Big(
  \partial_\mu \phi^a_{yk} \phi^b_{kl} \phi^c_{lx}
  - 2 \phi^a_{yk} \partial_\mu \phi^b_{kl} \phi^c_{lx}
  + \phi^a_{yk} \phi^b_{kl} \partial_\mu \phi^c_{lx}
  \Big).
\end{align}
The contractions in the calculation of the matrix element defined in 
Eq.~\eqref{eq:def_decay_const} give the loop integrals. 
Performing the loop integrals, we find the NLO current correction
to the decay constants,
\begin{align}
\label{eq:delta_f_curr}
 \delta f
 _{P_t^+} ^{\textrm{current}}
  \equiv -\frac{1}{6} \sum_{a} 
  \Bigg[ 
   \sum_{Q} l(Q_a)
  + 16\pi^2 \int \frac{d^4q}{(2\pi)^4} 
    \left( D_{xx}^a + D_{yy}^a - 2\theta^{at} D_{xy}^a \right)
  \Bigg].
\end{align}
Note that $\delta f_{P_t^+}^{\textrm{current}}$ is proportional to
$\delta f_{P_t^+}^Z$, which was shown for the taste Goldstone 
case in Ref.~\cite{Aubin:2003uc}.

Now we consider the NLO analytic contributions to the decay constants.
The terms from the NLO Lagrangian noted in Sec.~\ref{sec:review} 
(including the $\mathcal{O}(p^2 a^2)$ source operators) give 
the analytic contributions.
The contributions of $\mathcal{O}(p^2 a^2)$ terms may be written as 
$fa^2\mathcal{F}_t$.
Here $\mathcal{F}_t$ are linear combinations of the LECs of the Lagrangian, 
which are degenerate within the irreps of the lattice symmetry group.
As commented in Ref.~\cite{Sharpe:2004is}, there are no relations
between the SO(4)-violations in the pion masses and the SO(4)-violations
in the axial current decay constants, due to the contributions from the 
$\mathcal{O}(p^2 a^2)$
source operators.

The terms in Gasser-Leutwyler Lagrangian given in Eq.~\eqref{eq:GLterms}
contribute to the decay constants through wavefunction
renormalization and the current.
The wavefunction renormalization correction of the Gasser-Leutwyler terms can be 
calculated from the self-energy \cite{SWME:2011aa}.
Collecting all the NLO analytic corrections to the decay constants, we find
\begin{align}
\label{eq:delta_f_anal}
 \delta f_{P_t^+}^{\textrm{anal}} 
  = \frac{64}{f^2} L_4 \mu (m_u + m_d + m_s) 
   + \frac{8}{f^2} L_5 \mu (m_x + m_y) 
      + a^2 \mathcal{F}_t.
\end{align}
%

\section{\label{sec:results}Results}
The results given in Eqs.~\eqref{eq:delta_f_z}, \eqref{eq:delta_f_curr} and 
\eqref{eq:delta_f_anal} are the results in the 4+4+4 theory.
In order to formulate the results in the 1+1+1 theory (rooted staggered
chiral perturbation theory), we use the replica method 
\cite{Bernard:1993sv, Damgaard:2000gh, Bernard:2007ma}.
Applying the replica method to Eqs.~\eqref{eq:delta_f_z}, \eqref{eq:delta_f_curr} 
and \eqref{eq:delta_f_anal}, we find
\begin{align}
\label{eq:delta_f_rooted}
 \delta f_{P_F^+}
  &= \delta f_{P_F^+}^{\textrm{con}} + \delta f_{P_F^+}^{\textrm{disc}}, \\
 \label{eq:delta_f_anal_rooted}
 \delta f_{P_t^+}^{\textrm{anal}} 
  &= \frac{16}{f^2} L_4 \mu (m_u + m_d + m_s)
   + \frac{8}{f^2} L_5 \mu (m_x + m_y) 
      + a^2 \mathcal{F}_t.
\end{align}
where
\begin{align}
 \label{eq:delta_f_con}
 \delta f_{P_F^+}^{\textrm{con}}
  &\equiv -\frac{1}{32} \sum_{Q,B} g_B ~ l(Q_B), \\
 \label{eq:delta_f_disc}
 \delta f_{P_F^+}^{\textrm{disc}}
  &\equiv -2\pi^2 \int \frac{d^4q}{(2\pi)^4} 
    \Big( 
     D_{xx}^I + D_{yy}^I - 2 D_{xy}^I \nonumber \\
  & \qquad \qquad + 4 D_{xx}^V + 4 D_{yy}^V - 2\Theta^{VF} D_{xy}^V 
   + 4 D_{xx}^A + 4 D_{yy}^A - 2\Theta^{AF} D_{xy}^A
    \Big)
\,.
\end{align}
Here we performed the summation over $a$ within each taste SO(4) irrep for 
Eqs.~\eqref{eq:delta_f_z} and \eqref{eq:delta_f_curr},
$B$ and $F$ represent the taste $SO(4)_T$ irreps,
\begin{equation}
 B, F \in \{ I, V, T, A, P \},
\end{equation}
$t \in F$ and
\begin{equation}
 \Theta^{BF} \equiv \sum_{a \in B} \theta^{at}, \quad
 g_B \equiv \sum_{a \in B} 1.
\label{eq:coeff}
\end{equation}
The coefficients $\Theta^{BF}$ are given in Table~\ref{tab:coeff}.
\begin{table*}[t]
\caption{\label{tab:coeff}The coefficient $\Theta^{BF}$ defined in 
  Eq.~\protect\eqref{eq:coeff} is in row $B$ and column $F$.}
\begin{center}
\begin{tabular}{c|rrrrr}
\hline\hline
$B \backslash F$ & $V$ & $A$ & $T$ & $P$ & $I$ \\
\hline
$V$ & $-2$ & $2$ & $0$ & $-4$ & $\phantom{-}4$ \\
$A$ & $2$ & $-2$ & $0$ & $-4$ & $4$ \\
$T$ & $0$ &  $0$ & $-2$ & $6$ & $6$ \\
$P$ & $-1$ & $-1$ & $1$ & $1$ & $1$ \\
$I$ &  $1$ &  $1$ & $1$ & $1$ & $1$ \\
\hline\hline
\end{tabular}
\end{center}
\end{table*}
The superscripts \emph{con} and \emph{disc} in $\delta f_{P_F^+}^{\textrm{con}}$
and $\delta f_{P_F^+}^{\textrm{disc}}$ represent connected and disconnected
quark-flow contributions, respectively \cite{Aubin:2003uc}.

First, we consider partially quenched results for 1+1+1 and 2+1 flavor
cases in SU(3) SChPT.
The connected contributions to the decay constants in the partially quenched
1+1+1 flavor case are the same as the Eq.~\eqref{eq:delta_f_con}.
The disconnected contributions for the partially quenched 1+1+1 flavor case
are obtained by performing the integrals in Eq.~\eqref{eq:delta_f_disc} 
keeping all quark masses distinct,
\begin{align}
 \label{eq:delta_f_disc_su3_pq_a}
 \delta f
 &_{P_F^+, m_x \neq m_y}^{\mathrm{disc}} = \sum_{Z} \Bigg[
  \frac{1}{6} \Big\{ 
  D^{UDS}_{X \pi^0 \eta, X}(Z_I) l(Z_I)
  + D^{UDS}_{Y \pi^0 \eta, Y}(Z_I) l(Z_I)
  - 2 R^{UDS}_{XY \pi^0 \eta}(Z_I) l(Z_I) 
  \Big\} \nonumber \\
 &\quad + \frac{1}{4} a^2 \delta_V' \Big\{ 
  2D^{UDS}_{X \pi^0 \eta \eta', X}(Z_V) l(Z_V)
  + 2D^{UDS}_{Y \pi^0 \eta \eta', Y}(Z_V) l(Z_V)
 - \Theta^{VF} R^{UDS}_{XY \pi^0 \eta \eta'}(Z_V) l(Z_V) 
  \Big\} \nonumber \\
 &\quad + (V \rightarrow A)
 \Bigg] 
   + \frac{1}{6} \Big\{ 
    R^{UDS}_{X \pi^0 \eta}(X_I) \tilde{l}(X_I) 
  + R^{UDS}_{Y \pi^0 \eta}(Y_I) \tilde{l}(Y_I)
  \Big\} \nonumber \\
 &\qquad \qquad \qquad + \frac{1}{2} a^2 \delta_V' \Big\{
    R^{UDS}_{X \pi^0 \eta \eta'}(X_V) \tilde{l}(X_V)
  + R^{UDS}_{Y \pi^0 \eta \eta'}(Y_V) \tilde{l}(Y_V)
  \Big\} 
  + (V \rightarrow A).
\end{align}
For $m_x = m_y$, we find
\begin{align}
 \label{eq:delta_f_disc_su3_pq_b}
 \delta f
 &_{P_F^+, m_x = m_y}^{\mathrm{disc}} = 
 \frac{1}{4} a^2 \delta_V' (4-\Theta^{VF}) \Bigg[
   R^{UDS}_{X \pi^0 \eta \eta'}(X_V) \tilde{l}(X_V)
  + \sum_Z D^{UDS}_{X \pi^0 \eta \eta', X}(Z_V) l(Z_V)
 \Bigg] 
 +(V \rightarrow A).
\end{align}

For the 2+1 flavor case, the connected contributions are obtained by 
setting $m_u = m_d$ in Eq.~\eqref{eq:delta_f_con}, and
the disconnected contributions are obtained by setting $m_u = m_d$ and performing 
the integrals in Eq.~\eqref{eq:delta_f_disc}.
For $m_x \neq m_y$, we find
\begin{align}
 \label{eq:delta_f_xney_pq_21}
 \delta f
 &_{P_F^+, m_x \neq m_y}^{\mathrm{disc}} = \sum_{Z} \Bigg[
  \frac{1}{6} \Big\{ 
  D^{\pi S}_{X \eta, X}(Z_I) l(Z_I)
 + D^{\pi S}_{Y \eta, Y}(Z_I) l(Z_I)
  - 2 R^{\pi S}_{XY \eta}(Z_I) l(Z_I) 
  \Big\} \nonumber \\
 & + \frac{1}{4} a^2 \delta_V' \Big\{ 
  2D^{\pi S}_{X \eta \eta', X}(Z_V) l(Z_V)
  + 2D^{\pi S}_{Y \eta \eta', Y}(Z_V) l(Z_V)
  - \Theta^{VF} R^{\pi S}_{XY \eta \eta'}(Z_V) l(Z_V) 
  \Big\} \nonumber \\ 
 & + (V \rightarrow A)
 \Bigg] 
  + \frac{1}{6} \Big\{ 
    R^{\pi S}_{X \eta}(X_I) \tilde{l}(X_I) 
  + R^{\pi S}_{Y \eta}(Y_I) \tilde{l}(Y_I)
  \Big\} \nonumber \\
 &\qquad \qquad \quad + \frac{1}{2} a^2 \delta_V' \Big\{
    R^{\pi S}_{X \eta \eta'}(X_V) \tilde{l}(X_V)
  + R^{\pi S}_{Y \eta \eta'}(Y_V) \tilde{l}(Y_V)
  \Big\} 
  + (V \rightarrow A).
\end{align}
For $m_x = m_y$, we find
\begin{align}
 \label{eq:delta_f_xeqy_pq_21}
 \delta f
 &_{P_F^+, m_x = m_y}^{\mathrm{disc}} = 
 \frac{1}{4} a^2 \delta_V (4-\Theta^{VF}) \Bigg[
   R^{\pi S}_{X \eta \eta'}(X_V) \tilde{l}(X_V)
  + \sum_Z D^{\pi S}_{X \eta \eta', X}(Z_V) l(Z_V)
 \Bigg] 
 +(V \rightarrow A).
\end{align}

Next, we consider the partially quenched results for the 1+1+1 flavor 
case in SU(2) SChPT.
The connected contributions to the decay constants are obtained by
dropping terms corresponding to strange sea quark loops
from Eq.~\eqref{eq:delta_f_con}.
The disconnected contributions are obtained from
Eqs.~\eqref{eq:delta_f_disc_su3_pq_a},
\eqref{eq:delta_f_disc_su3_pq_b} and \eqref{eq:delta_f_disc},
by taking the SU(2) limit treating $x$ and $y$ as light quarks 
($m_x, m_y, m_u, m_d \ll m_s$),
\begin{align}
 \label{eq:delta_f_xney_su2_pq}
 \delta f
 &_{P_F^+, m_x \neq m_y}^{\mathrm{disc}} = \sum_{Z} \Bigg[
  \frac{1}{4} \Big\{ 
  D^{UD}_{X \pi^0, X}(Z_I) l(Z_I)
 + D^{UD}_{Y \pi^0, Y}(Z_I) l(Z_I)
  - 2 R^{UD}_{XY \pi^0}(Z_I) l(Z_I) 
  \Big\} \nonumber \\
 &\quad + \frac{1}{4} a^2 \delta_V' \Big\{ 
  2D^{UD}_{X \pi^0 \eta, X}(Z_V) l(Z_V)
  + 2D^{UD}_{Y \pi^0 \eta, Y}(Z_V) l(Z_V)
  - \Theta^{VF} R^{UD}_{XY \pi^0 \eta}(Z_V) l(Z_V) 
  \Big\} + (V \rightarrow A)
 \Bigg] \nonumber \\
 & + \frac{1}{4} \Big\{ 
    R^{UD}_{X \pi^0}(X_I) \tilde{l}(X_I) 
  + R^{UD}_{Y \pi^0}(Y_I) \tilde{l}(Y_I)
  \Big\}  \nonumber \\
 & + \frac{1}{2} a^2 \delta_V' \Big\{
    R^{UD}_{X \pi^0 \eta}(X_V) \tilde{l}(X_V)
  + R^{UD}_{Y \pi^0 \eta}(Y_V) \tilde{l}(Y_V)
  \Big\} 
  + (V \rightarrow A),
\end{align}
and
\begin{align}
 \label{eq:delta_f_xeqy_su2_pq}
 \delta f
 &_{P_F^+, m_x = m_y}^{\mathrm{disc}} = 
 \frac{1}{4} a^2 \delta_V' (4-\Theta^{VF}) \Bigg[
   R^{UD}_{X \pi^0 \eta}(X_V) \tilde{l}(X_V)
  + \sum_Z D^{UD}_{X \pi^0 \eta, X}(Z_V) l(Z_V)
 \Bigg] 
 +(V \rightarrow A).
\end{align}

For the 2+1 flavor case in SU(2) SChPT, the connected contributions are obtained by 
setting $m_u = m_d$ for the 1+1+1 case, and
the disconnected contributions are obtained by setting $m_u = m_d$
in Eqs.~\eqref{eq:delta_f_disc_su3_pq_a} and \eqref{eq:delta_f_disc_su3_pq_b},
\begin{align}
 \delta f
 &_{P_F, m_x \neq m_y}^{\mathrm{disc}} = \sum_{Z} \Bigg[
  -\frac{1}{2}
  R^{\pi}_{XY}(Z_I) l(Z_I) 
  \nonumber \\
 &\quad + \frac{1}{4} a^2 \delta_V' \Big\{ 
  2D^{\pi}_{X \eta, X}(Z_V) l(Z_V)
  + 2D^{\pi}_{Y \eta, Y}(Z_V) l(Z_V)
 - \Theta^{VF} R^{\pi}_{XY \eta}(Z_V) l(Z_V) 
  \Big\} 
  + (V \rightarrow A)
 \Bigg] \nonumber \\
 & + \frac{1}{4} \Big\{ 
    l(X_I) + (\pi_I-X_I)\tilde{l}(X_I) 
  + l(Y_I) + (\pi_I-Y_I)\tilde{l}(Y_I)
  \Big\} \nonumber \\
 & + \frac{1}{2} a^2 \delta_V' \Big\{
    R^{\pi}_{X \eta}(X_V) \tilde{l}(X_V)
  + R^{\pi}_{Y \eta}(Y_V) \tilde{l}(Y_V)
  \Big\} 
  + (V \rightarrow A),
\end{align}
and
\begin{align}
\label{eq:delta_f_xney_su2_pq_21}
 \delta f
 &_{P_F, m_x = m_y}^{\mathrm{disc}} = 
 \frac{1}{4} a^2 \delta_V' (4-\Theta^{VF}) \Bigg[
   R^{\pi}_{X \eta}(X_V) \tilde{l}(X_V)
  + \sum_Z D^{\pi}_{X \eta, X}(Z_V) l(Z_V)
 \Bigg] 
 +(V \rightarrow A).
\end{align}
%

\section{\label{sec:conclusion}Conclusion}
In Eqs.~\eqref{eq:delta_f_disc_su3_pq_a} -- \eqref{eq:delta_f_xeqy_pq_21},
we present the NLO corrections to the pion and kaon decay constants
for the partially quenched case calculated in the SU(3) SChPT;
in Eqs.~\eqref{eq:delta_f_xney_su2_pq} -- \eqref{eq:delta_f_xney_su2_pq_21},
we present the NLO corrections to the pion and kaon decay constants
for the partially quenched case calculated in the SU(2) SChPT.
As one can see in Eqs.~\eqref{eq:delta_f_disc} and \eqref{eq:delta_f_anal_rooted},
the only differences between taste Goldstone and taste non-Goldstone cases
are the $\Theta^{BF}$ factors multiplying $D_{xy}^{A,V}$ and the generalized 
constants $\mathcal{F}_t$ in the analytic contribution. 
$\Theta^{BF}$ originates from the trace of taste generators and affects
only the flavor-charged disconnected propagator, $D_{xy}^{A,V}$.
$\mathcal{F}_t$ are degenerate within the lattice symmetry group, and 
there are no relations between the SO(4)-violations in the pion masses 
and the SO(4)-violations in the axial current decay constants. 
Using these results, it is possible to improve determinations of the
decay constants, quark masses and the Gasser-Leutwyler constants
by analyzing lattice data from taste non-Goldstone channels.

\section{Acknowledgments}
W.~Lee is supported by the Creative Research
Initiatives Program (2012-0000241) of the NRF grant funded by the
Korean government (MEST), and acknowledges
support from KISTI supercomputing
center through the strategic support program for the supercomputing
application research [No. KSC-2011-G2-06].

\end{document}